\documentclass[a4paper,11pt]{article}
\usepackage{pos}
\usepackage{graphicx}
\usepackage{amsmath,amsthm,bm,bigdelim,multirow}
\usepackage{color}
\usepackage{booktabs}
\usepackage{hyperref}
\usepackage{dsfont}
\usepackage{here}
\usepackage{subcaption}

\newcommand{\kslash}{k\kern-1ex /}
\newcommand{\pslash}{p\kern-1ex /}
\newcommand{\qslash}{q\kern-1ex /}
\newcommand{\lslash}{l\kern-1ex /}
\newcommand{\sslash}{s\kern-1ex /}
\newcommand{\Dslash}{D\kern-1.2ex /}

\newcommand{\beqa}{\begin{eqnarray}}
\newcommand{\eeqa}{\end{eqnarray}}
\newcommand{\Tr}{{\rm Tr}}
\newcommand{\be}{\[}
\newcommand{\ee}{\]}
\newcommand{\bd}{\begin{description}}
\newcommand{\ed}{\end{description}}

\newcommand{\ben}{\begin{eqnarray}}
\newcommand{\een}{\end{eqnarray}}

\def\lsim{\raise0.3ex\hbox{$<$\kern-0.75em\raise-1.1ex\hbox{$\sim$}}}
\def\gsim{\raise0.3ex\hbox{$>$\kern-0.75em\raise-1.1ex\hbox{$\sim$}}}
\def\simgt{\rlap{\lower 3.5 pt\hbox{$\mathchar \sim$}}\raise 2.0pt \hbox {$>$}}
\def\simlt{\rlap{\lower 3.5 pt\hbox{$\mathchar \sim$}}\raise 2.0pt \hbox {$<$}}

\title{Tensor renormalization group study of (1+1)-dimensional O(3) nonlinear sigma model with and without finite chemical potential}
\ShortTitle{TRG study for (1+1)$d$ O(3) NLSM w/ and w/o finite chemical potential}

\author[a]{Xiao Luo}
\author*[b]{Yoshinobu Kuramashi}

\affiliation[a]{Graduate School of Pure and Applied Sciences, University of Tsukuba, Ibaraki 305-8571, Japan}

\affiliation[b]{Center for Computational Sciences, University of Tsukuba, Ibaraki 305-8577, Japan}

\emailAdd{luo@het.ph.tsukuba.ac.jp}
\emailAdd{kuramasi@het.ph.tsukuba.ac.jp}

\abstract{We study (1+1)-dimensional O(3) nonlinear sigma model using the tensor renormalization group method with the infinite limit of the bond dimension $D_{\rm cut}\rightarrow \infty$.
At the vanishing chemical potential $\mu=0$,  we investigate the von Neumann and  R\'enyi types of entanglement entropies. The central charge is determined to be $c=1.97(9)$ by using the asymptotic scaling properties of the entropies. We also examine the consistency between two entropies.  In the finite density region with $\mu\ne 0$, where this model suffers from the sign problem in the standard Monte Carlo approach, we investigate the properties of the quantum phase transition. We determine the transition point $\mu_{\rm c}$ and the critical exponent of the correlation length $\nu$ from the $\mu$ dependence of the number density in the thermodynamic limit. The dynamical critical exponent $z$ is also extracted from the scaling behavior of the temporal correlation length as a function of $\mu$. This is the first successful calculation of the dynamical critical exponent with the TRG method.}

\FullConference{The 41st International Symposium on Lattice Field Theory (LATTICE2024)\\
 28 July - 3 August 2024\\
Liverpool, UK\\}

\begin{document}
\maketitle

\section{Introduction}
\label{sec:intro}

The basic idea of the tensor renormalization group (TRG) method~\footnote{In this paper, the ``TRG method" or the ``TRG approach" refers to not only the original numerical algorithm proposed by Levin and Nave \cite{Levin:2006jai} but also its extensions~\cite{PhysRevB.86.045139,Shimizu:2014uva,PhysRevLett.115.180405,Sakai:2017jwp,PhysRevLett.118.110504,Hauru:2017tne,Adachi:2019paf,Kadoh:2019kqk,Akiyama:2020soe,PhysRevB.105.L060402,Akiyama:2022pse}.} was originally proposed in the field of condensed matter physics in 2007~\cite{Levin:2006jai}. In the past decade the TRG method has been getting applied to the particle physics. Although the inital research target was focused on the phase transitions of two-dimensional (2$d$) models, recent studies cover those for 4$d$ models with the scalar, gauge and fermion fields~\cite{Akiyama:2019xzy,Akiyama:2020ntf,Akiyama:2021zhf,Akiyama:2020soe,Akiyama:2022eip,Akiyama:2023hvt}. The particle physicists are attracted by the following characteristc features in the TRG method: (i) no sign problem, (ii) logarithmic computational cost on the system size, (iii) direct manipulation of the Grassmann variables, (iv) evaluation of the partition function or the path-integral itself. So far much attention has been paid to the featute (i)~\cite{Shimizu:2014uva,Shimizu:2014fsa,Kawauchi:2016xng,Kawauchi:2016dcg,Yang:2015rra,Shimizu:2017onf,Takeda:2014vwa,Kadoh:2018hqq,Kadoh:2019ube,Kuramashi:2019cgs,Akiyama:2020ntf,Akiyama:2020soe,Akiyama:2021xxr,Akiyama:2021glo,Nakayama:2021iyp,Akiyama:2023hvt,Luo:2022eje,Akiyama:2024qer},

In this report we investigate the (1+1)$d$ O(3) nonlinear sigma model (O(3) NLSM) with and without finite chemical potential. This model is massive and shares the property of asymptotic freedom with the (3+1)$d$ non-Abelian gauge theories so that it should be a good testbed before exploring to investigate the properties of QCD. At $\mu=0$ we measure the von Neumann and R\'enyi types of entanglement entropies taking advantage of the above feature (iv)~\cite{Luo:2023ont}. The central charge is determined from the asymptotic scaling properties of the entanglement entropies. We also make a direct comparison of both entropies and discuss the consistency between them.  At $\mu\ne 0$ we perform a detailed study of the quantum phase transition, which is achieved thanks to the above features (i) and (ii)~\cite{Luo:2024lbh}. We determine the transition point $\mu_{\rm c}$, the critical exponent $\nu$ and the dynamical critical exponent $z$, where $\nu=0.5$ and $z=2$ are theoretically expected based on the equivalence between the (1+1)$d$ O(3) NLSM at finite density and the integer-spin Heisenberg chain with a magnetic field~\cite{Dzhaparidze_1978,Pokrovsky_1979,Schulz_1980,Schulz_1986,Affleck_1990}.

This report is organized as follows. In Sec.~\ref{sec:formulation}, we define the action of the (1+1)$d$ O(3) NLSM with finite $\mu$ on the lattice and give the tensor network representation. 
In Sec.~\ref{sec:results} we present the numerical results.
Section~\ref{sec:summary} is devoted to summary.

\section{Formulation}
\label{sec:formulation}
\subsection{Tensor network representation}
\label{subsec:tensornetrep}

We consider the partition function of the O(3) NLSM with the chemical potential $\mu$ on a (1+1)$d$ lattice $\Lambda_{1+1}=\{(n_1,n_2)\ \vert n_1=1,\dots,L, n_2=1,\dots,N_t\}$ whose volume is $V=L\times N_t$. The temperature $T$ is given by $T=1/N_t$. We set the lattice spacing $a=1$ unless necessary. A real three-component unit vector $\boldmath{s}(n)$ resides on the sites $n$ and satisfies the periodic boundary conditions $\boldmath{s}(n+{\hat \nu}L)=\boldmath{s}(n)$ ($\nu=1,2$). The lattice action $S$ is defined as
\begin{equation}
  S = -\beta \sum_{n\in\Lambda_{1+1},\nu} \sum_{\lambda,\gamma=1}^{3} s_\lambda(\Omega_n) D_{\lambda\gamma}(\mu,\hat{\nu}) s_\gamma(\Omega_{n+\hat{\nu}}),
  \label{eq:action}
\end{equation}
where the spin $s(\Omega)$ and matrix $D(\mu,\hat{\nu})$ are expressed as
\begin{align}
	& s(\Omega)= \left(\begin{array}{l}\cos\theta\\
		\sin\theta\cos\phi\\
		\sin\theta\sin\phi
	\end{array}\right), \\
	& D(\mu,\hat{\nu}) = \left(\begin{array}{rrr} 1 & ~ & ~ \\
		~ & \cosh(\delta_{2,\nu}\mu) & -i\sinh(\delta_{2,\nu}\mu) \\
		~ & i\sinh(\delta_{2,\nu}\mu )& \cosh(\delta_{2,\nu}\mu)
	\end{array}\right) 
\end{align}
with 
\begin{equation}
	\label{U:representation}
	\begin{array}{c}
		 \Omega=(\theta,\phi) \quad, ~\theta \in (0,\pi],~\phi \in (0,2\pi]. 
	\end{array}
\end{equation}
Note that we introduce the chemical potential to the rotation between the second and third components.




The partition function and its measure are written as
\begin{align}
	\label{eq:partitionfunction2}
	Z &= \int {\cal D}\Omega \prod_{n,\nu} e^{\beta  \sum_{\lambda,\gamma=1}^{3}s_\lambda(\Omega_n) D_{\lambda\gamma}(\mu,\hat{\nu}) s_\gamma(\Omega_{n+\hat{\nu}})}, \\
	{\cal D}\Omega &= \prod_{p=1}^{V} \frac{1}{4\pi} \sin(\theta_p) d\theta_p d\phi_p~.
\end{align}
We discretize the integration (\ref{eq:partitionfunction2}) with the Gauss-Legendre quadrature~\cite{Kuramashi:2019cgs} after changing the integration variables:
\ben
-1 \le \alpha&=&\frac{1}{\pi}\left(2\theta-\pi \right)\le 1, \\
-1 \le \beta&=&\frac{1}{\pi}\left(\phi-\pi \right)\le 1. 
\een
We obtain
\begin{equation}
	Z = \sum_{ \{\Omega_1\},\cdots,\{\Omega_V\}} \left( \prod_{n=1}^{V} \frac{\pi}{8}  \sin(\theta(\alpha_{a_n})) w_{a_n} w_{b_n} \right) \prod_{\nu} M_{\Omega_n,\Omega_{n+\hat{\nu}}}
\end{equation}
with $\Omega_n=(\theta(\alpha_{a_n}),\phi(\beta_{b_n}))\equiv (a_n,b_n)$, where $\alpha_{a_n}$ and $\beta_{b_n}$ are $a$- and $b$-th roots of the $K$-th Legendre polynomial $P_{K}(s)$ on the site $n$, respectively. $\sum_{ \{\Omega_n\}}$ denotes $\sum_{a_n=1}^{K}\sum_{b_n=1}^{K}$.
$M$ is a 4-legs tensor defined by
\begin{equation}
	M_{a_n,b_n,a_{n+\hat{\nu}}, b_{n+\hat{\nu}}}  =\exp\left\{ \beta \sum_{\lambda,\gamma=1}^{3}s_\lambda(a_n,b_n) D_{\lambda\gamma}(\mu,\hat{\nu}) s_\gamma (a_{n+\hat{\nu}}, b_{n+\hat{\nu}}) \right\}~.
\end{equation} 
The weight factor $w$ of the Gauss-Legendre quadrature is defined as
\begin{equation}
	w_{a_n} = \frac{2(1-{\alpha_{a_n}}^2)}{K^2P^2_{K-1}({\alpha_{a_n}})},\quad
	w_{b_n} = \frac{2(1-{\beta_{b_n}}^2)}{K^2P^2_{K-1}({\beta_{b_n}})}.
\end{equation}
Throughout this report we employ $K=100$.
After performing the singular value decomposition (SVD) on $M$, we obtain
\begin{equation}
	M_{a_n,b_n,a_{n+\hat{\nu}}, b_{n+\hat{\nu}}} \simeq \sum_{i_n=1}^{D_\text{cut}} U_{a_n,b_n, i_n} (\nu) \sigma_{i_n}(\nu) V^\dagger_{i_n,a_{n+\hat{\nu}}, b_{n+\hat{\nu}}} (\nu),
\end{equation}
where $U$ and $V$ denote unitary matrices and $\sigma$ is a diagonal matrix with the largest $D_{\text{cut}}$ singular values of $M$ in the descending order.
We can obtain the tensor network representation of the O(3) NLSM on the site $n\in\Lambda_{1+1}$
\begin{align}
	T_{x_n, x'_n, y_n, y'_n} &= \frac{\pi}{8} \sqrt{\sigma_{x_n}(1) \sigma_{x'_n}(1) \sigma_{y_n}(2) \sigma_{y'_n}(2) } \sum_{a_n, b_n} w_{a_n} w_{b_n} \sin(\theta_{a_n}) \nonumber \\
	&\quad \times V^\dagger_{x_n,a_n,b_n} (1) U_{a_n,b_n, x'_n} (1) V^\dagger_{y_n,a_n,b_n} (2) U_{a_n,b_n, y'_n} (2) .
\end{align} 
Here the bond dimension of tensor $T$ is given by  $D_{\text{cut}}$, which controls the numerical precision in the TRG method. The tensor network representation of the partition function is given by
\begin{equation}
  Z \simeq \sum_{x_0 x'_0 y_0 y'_0 \cdots} \prod_{n \in \Lambda_{1+1}} T_{x_n x'_n y_n y'_n} = \Tr \left[T \cdots T\right]~.
  \label{eq:Z_TN}
\end{equation}
In order to evaluate $Z$ we employ the higher order tensor renormalization group (HOTRG) algorithm~\cite{PhysRevB.86.045139}.

\subsection{Correlation length and entanglement entropies}
\label{subsec:clee}

We evaluate the temporal correlation length $\xi_t$ with 
\ben
\xi_t = {N_t \over \ln\left(\frac{\lambda_0}{\lambda_1}\right)}, 
\een
where $\lambda_0$ and $\lambda_1$ is the largest and the second largest eigenvalues of the density matrix $\rho_{yy^\prime}=\Tr_x T^*_{xxyy^\prime}$ with $T^*$ the reduced single tensor obtained by HOTRG.

For calculation of the entanglement entropies we consider the system consisting of two subsystems A nad B with the same size of $L\times N_t$. 
The von Neumann entropy is obtained by
\ben
S_A = -\Tr_A \rho_A \log(\rho_A)
\een
with $\rho_A\simeq {1\over Z}\Tr_B T^*_{xx'y_By_B}T^*_{x'xy_Ay'_A} = M_{y_A,y'_A}$, where $\Tr_B$ denotes the trace restricted to the subsystem B. On the other hand, the R{\'e}nyi entropy is defined by
\ben
S_A^{(n)} = \frac{\log \Tr_A \rho_A^n}{1-n},
\een
with $\rho_A^n$ the $n$th matrix power of $\rho_A$. Both entropies are related by
\ben
\lim_{n\rightarrow 1} S_A^{(n)} = S_A.
\een

\section{Numerical results}
\label{sec:results}
\subsection{Entanglement entropies with $\mu=0$}
\label{subsec:ee}

\begin{figure}[htbp]
	\centering
	\includegraphics[width=0.5\hsize]{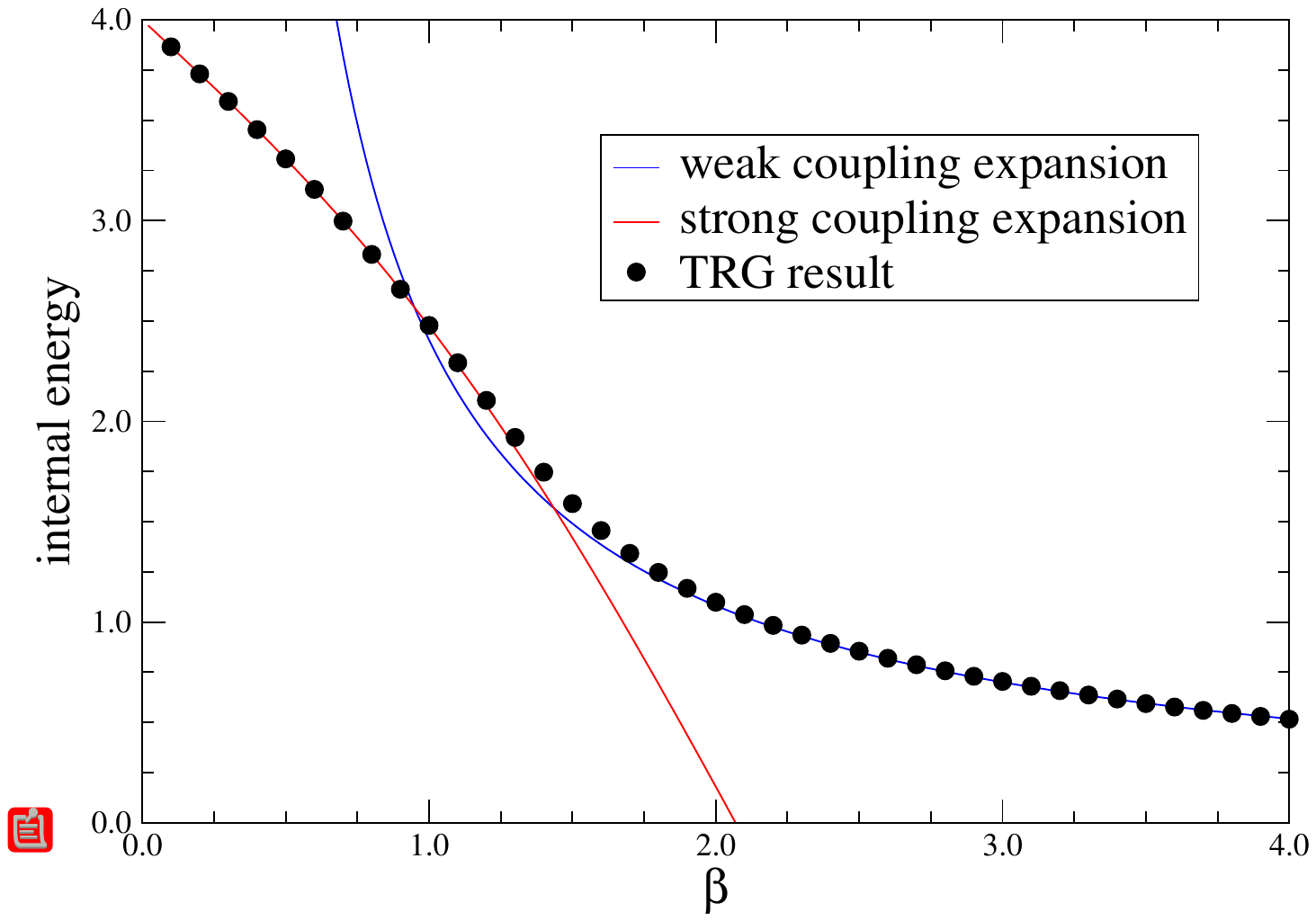}
	\caption{$\beta$ dependence of internal energy at $\mu=0$ on a $1024\times 1024$ lattice. Solid curves represent the results of the strong and weak coupling expansions.}
  	\label{fig:energy}
\end{figure}

Before discussing the entanglement entropies, it may be instructive to show the  results for the internal energy at $\mu=0$ obtained by the impure tensor method with $D_{\rm cut}=48$~\cite{Luo:2022eje}. In Fig.~\ref{fig:energy} we compare the TRG results with the strong and weak coupling expansions. In the strong coupling region we observe that our result show good consistency with the strong coupling expansion up to $\beta\sim 1.2$. On the other hand, the result starts to follow the weak coupling expansion curve around $\beta\sim 2.0$.

The density matrix $\rho_A$ is evaluated using HOTRG with the bond dimension $D_{\rm cut}\in [10,130]$.  We choose $\beta=1.4$, 1.5, 1.6 and 1.7 for the coupling constant to keep the condition $a\ll \xi \ll L$, where the correlation length $\xi$ was precisely measured in Ref.~\cite{Wolff:1989hv}: $\xi=6.90(1)$, 11.09(2), 19.07(6) and 34.57(7) at $\beta=1.4$, 1.5, 1.6 and 1.7, respectively.

\begin{figure}[htbp]
	\begin{minipage}[t]{0.48\hsize}
    		\centering
   	 	\includegraphics[width=1.0\hsize]{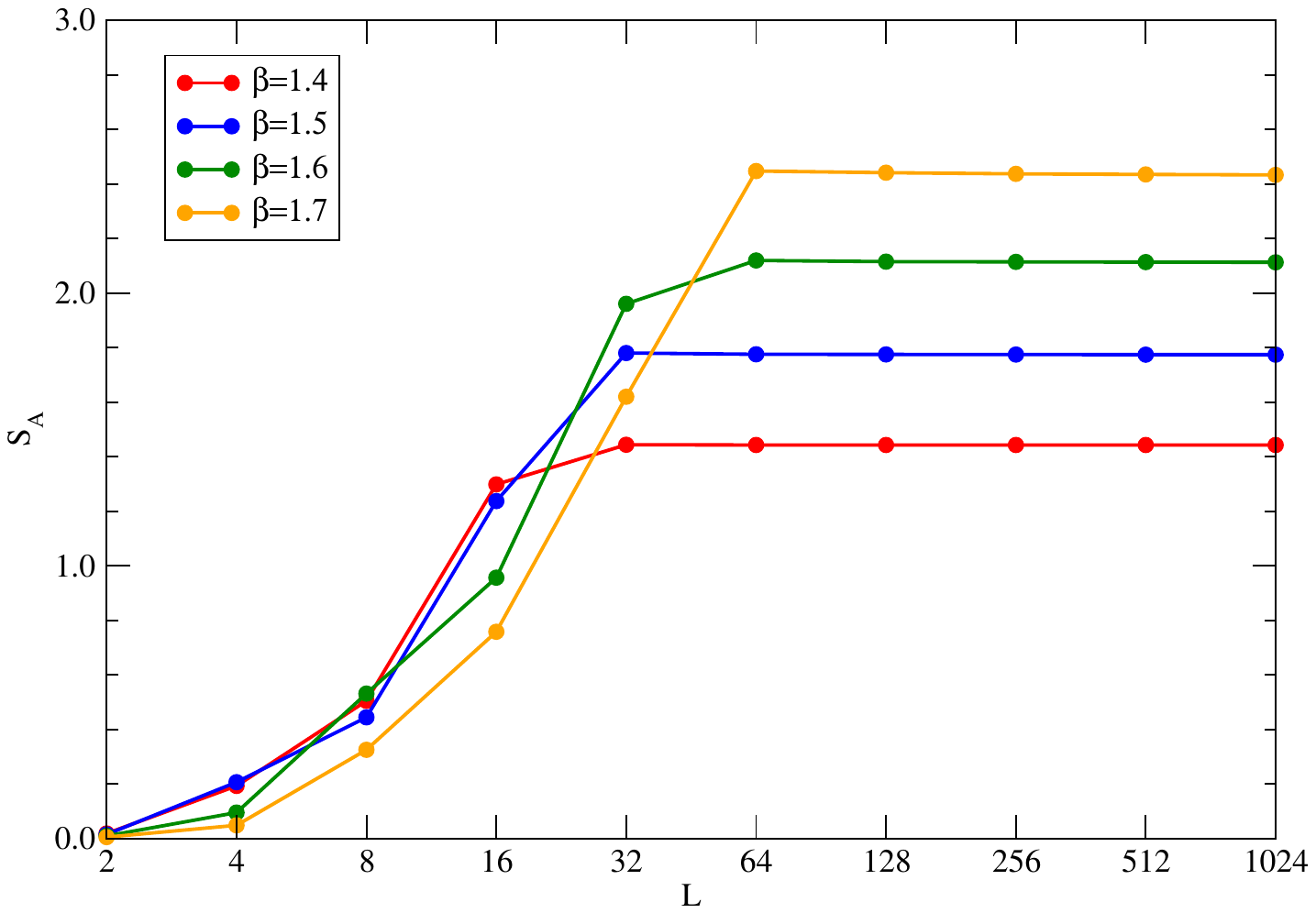}
  		\caption{$L$ dependence of von Neumann entropy $S_A$ at $N_t=1024$ in $1.4\le \beta \le 1.7$ with $D_{\rm cut}=130$.}
  		\label{fig:sa_beta}
  	\end{minipage}
  	\hspace*{3mm}
	\begin{minipage}[t]{0.48\hsize}
    		\centering
    		\includegraphics[width=1.0\hsize]{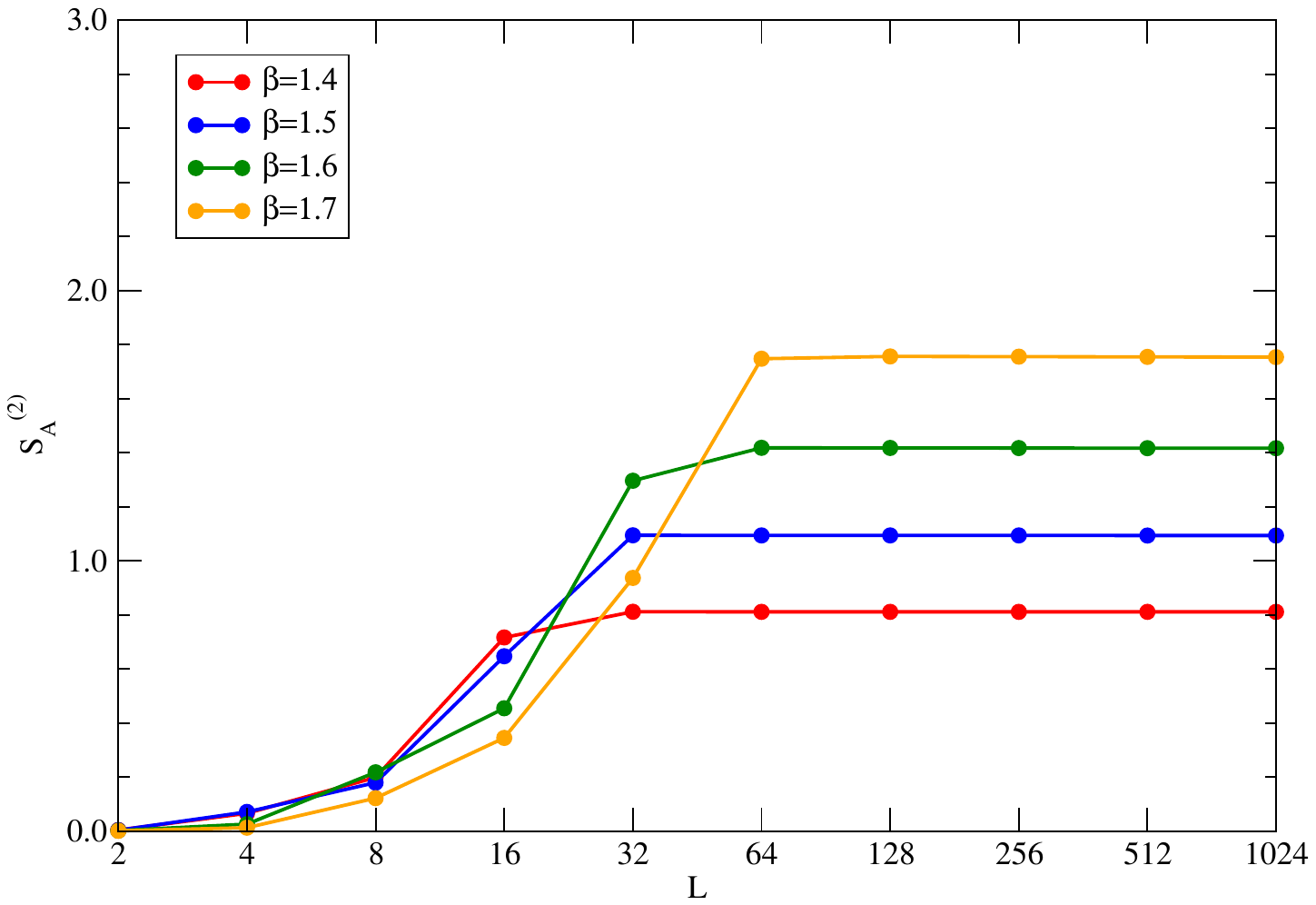}
  		\caption{$L$ dependence of 2nd-order R{\'e}nyi entropy $S_A^{(2)}$ at $N_t=1024$ in $1.4\le \beta \le 1.7$ with $D_{\rm cut}=130$.}
  		\label{fig:sa2_beta}
	\end{minipage}
\end{figure}

Figure~\ref{fig:sa_beta} plots $L$ dependence of the von Neumann entropy $S_A$ at $N_t=1024$ with $1.4\le \beta \le 1.7$, where $N_t=1024$ is large enough to be regarded as the zero temperature limit. We observe that $S_A(L)$ shows plateau behavior once the interval $L$ goes beyond the correlation length. 
As $\xi$ increases for larger $\beta$, the plateau of $S_A(L)$ starts at larger $L$ and its value is increased according to the theoretical expectation of $S_A(L)\sim \frac{c}{3}\ln \xi$ with $c$ the central charge under the condition of $\xi\ll L$~\cite{Calabrese:2004eu}. For a comparative purpose we also plot the $L$ dependence of the 2nd-order R{\'e}nyi entropy $S_A^{(2)}$ in Fig.~\ref{fig:sa2_beta}. Both entropies show similar behaviors, though the plateau values of $S_A^{(2)}$ are smaller than those of  $S_A$ according to the theoretical expectation $S_A^{(n)}(L)\sim \frac{c}{6}(1+1/n)\ln \xi$~\cite{Calabrese:2004eu}.

\begin{figure}[htbp]
	\begin{minipage}[t]{0.48\hsize}
    		\centering
   	 	\includegraphics[width=1.0\hsize]{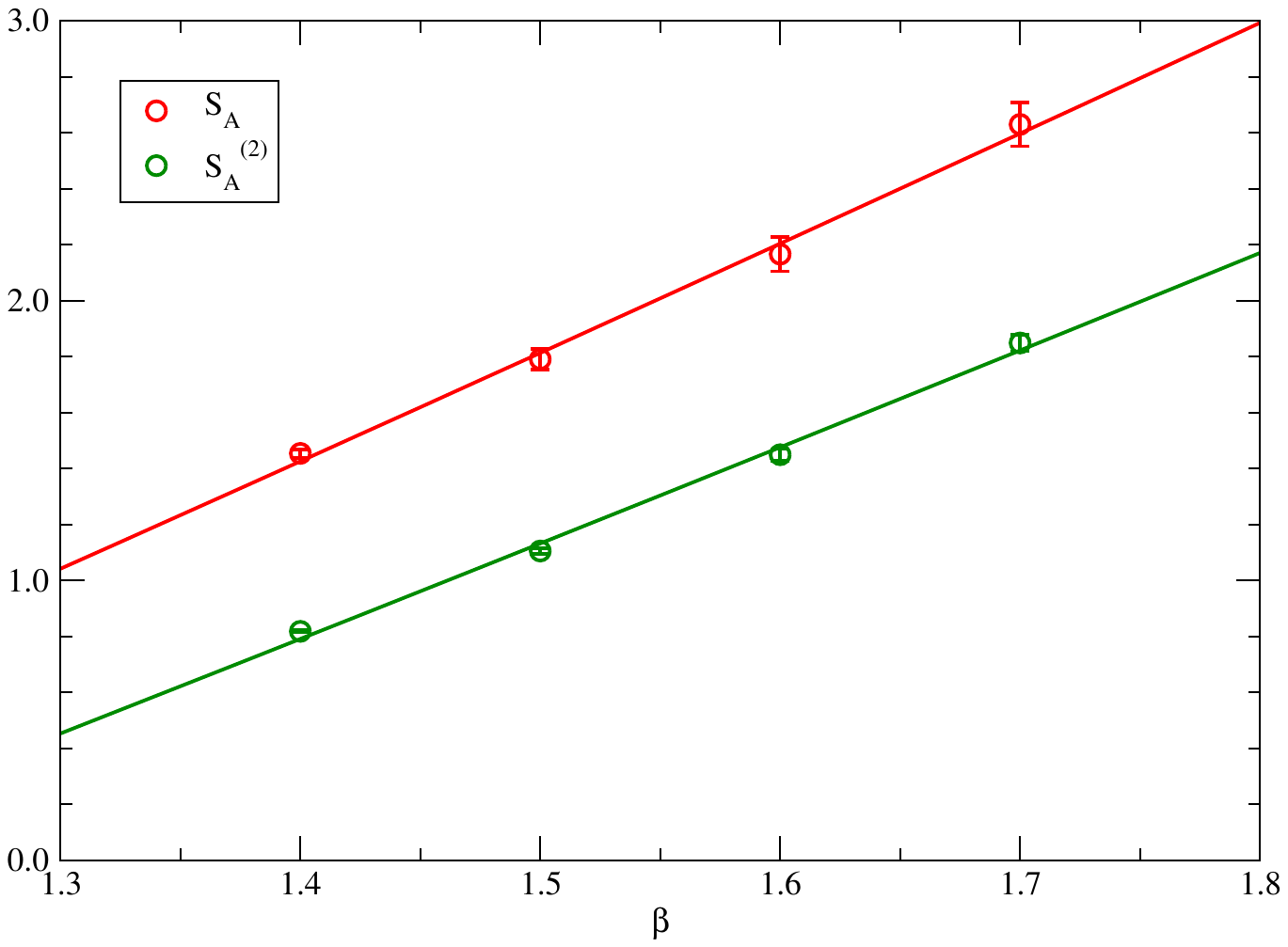}
  		\caption{$\beta$ dependence of von Neumann entropy $S_A$ at $(L,N_t)=(128,1024)$ in $1.4\le \beta \le 1.7$ with $D_{\rm cut}=130$.}
  		\label{fig:s_beta}
  	\end{minipage}
  	\hspace*{3mm}
	\begin{minipage}[t]{0.48\hsize}
    		\centering
    		\includegraphics[width=1.0\hsize]{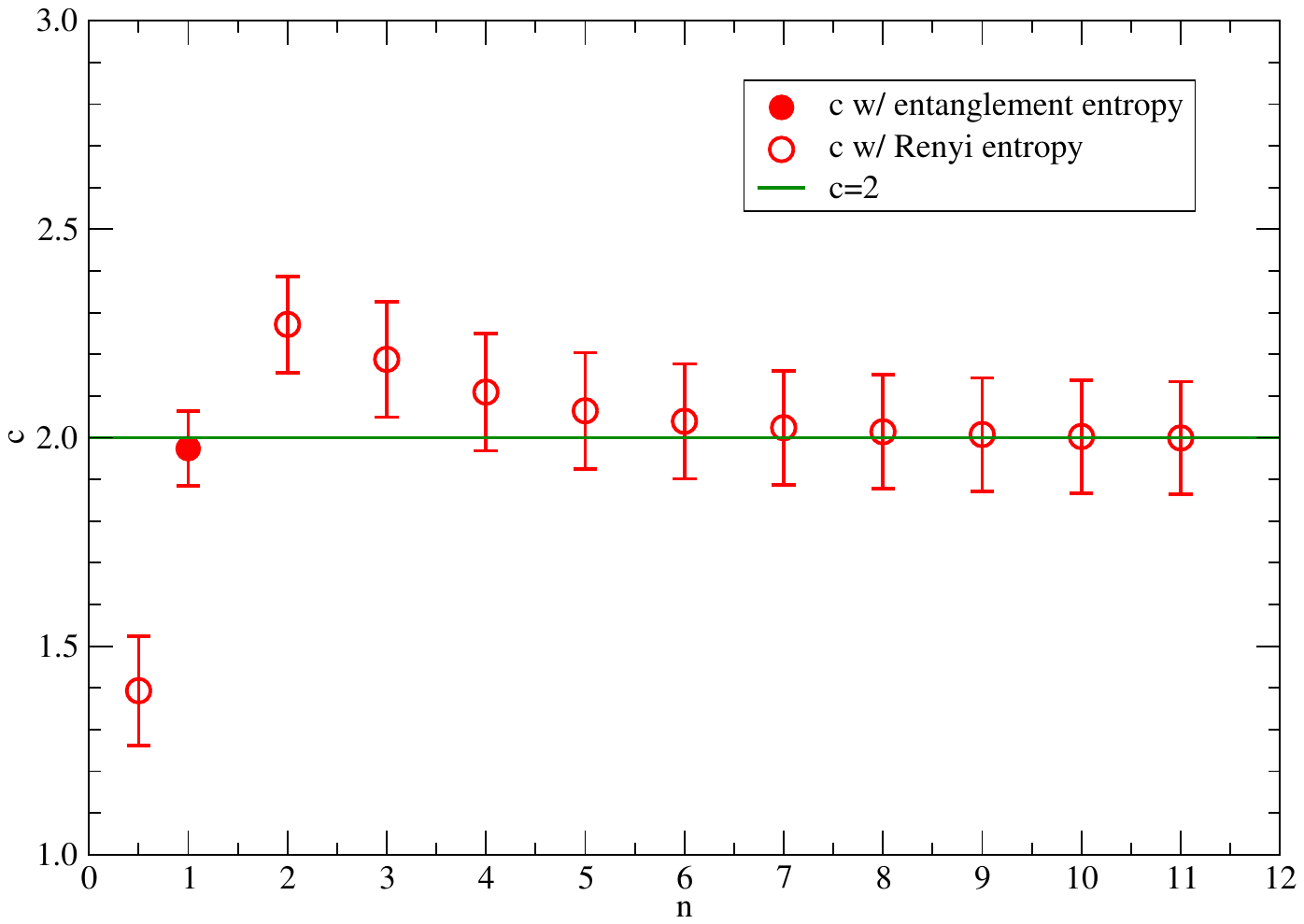}
  		\caption{$n$ dependence of $n$th-order R{\'e}nyi entropy $S_A^{(n)}$ at $(L,N_t)=(128,1024)$ in $1.4\le \beta \le 1.7$ with $D_{\rm cut}=130$.}
  		\label{fig:cre}
	\end{minipage}
\end{figure}

In Fig.~\ref{fig:s_beta} we show the $\beta$ dependence of $S_A(L=128)$ and $S_A^{(2)}(L=128)$ at $N_t=1024$, which are obtained by linear extrapolations in terms of $1/D_{\rm cut}$ to remove the finite $D_{\rm cut}$ effects. We extract the central charge $c$ by fitting the data with the following functions:
\ben
S_A&=&\frac{c}{3}\left(2\pi\beta-\ln\beta \right)+{\rm const.}
\label{eq:sa_beta}, \\
S_A^{(n)}&=&\frac{c}{6}\left(1+\frac{1}{n}\right)\left(2\pi\beta-\ln\beta \right)+{\rm const.},
\label{eq:san_beta}
\een
with $n=2$, where we use the perturbative $\beta$ dependence of $\xi\propto 1/\beta \exp(2\pi\beta)$.  For the von Neumann entropy we obtain $c=1.97(9)$, which is consistent with $c=2.04(14)$ obtained by the MPS method~\cite{Bruckmann:2018usp}. On the other hand, the value of $c=2.27(16)$ extracted from the 2nd-order R{\'e}nyi entropy is slightly larger than that from the von Neumann entropy. Repeating the same calculation for other $n$th-order R{\'e}nyi entropy we obtain the $n$ dependence of the central charge $c$ shown in Fig.~\ref{fig:cre}. As $n$ increases the central charge seems to converges to $c=2$ and becomes consistent with $c=1.97(9)$ determined from the von Neumann entropy. This convergence behavior may be explained by the fact that the largest eigenvalue in the density matrix, which is most presisely calculated, gives dominant contribution to the R{\'e}nyi entropy as $n$ increases.

\begin{figure}[htbp]
	\centering
	\includegraphics[width=0.5\hsize]{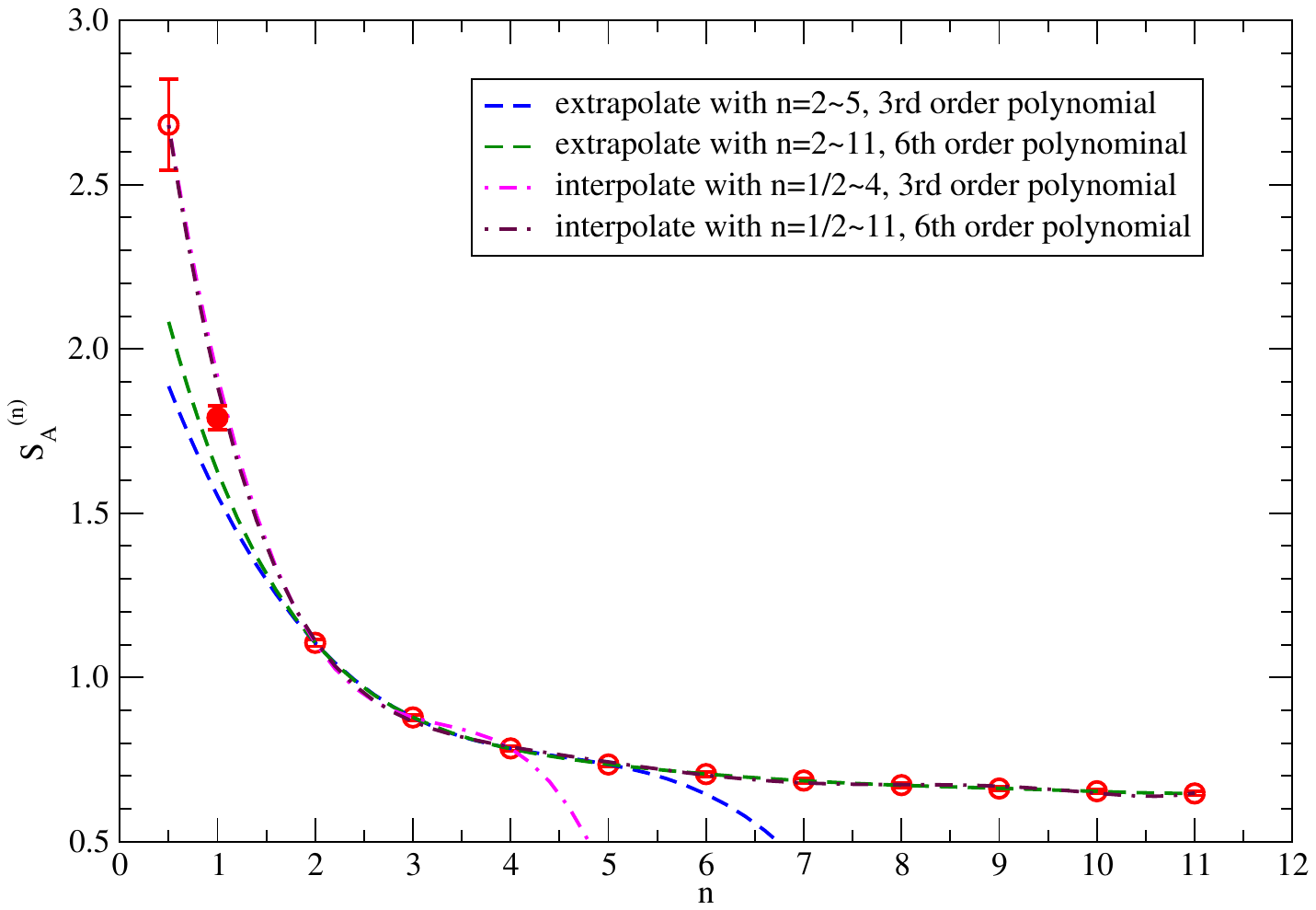}
	\caption{$n$ dependence of $n$th-order R\'enyi entropy with $N_t=1024$ at $\beta=1.5$. Solid symbol at $n=1$ denotes the von Neumann entropy. All the results are extrapolated values at $D_{\rm cut}\rightarrow \infty$.}
  	\label{fig:renyi_n}
\end{figure}

In Fig.~\ref{fig:renyi_n} we plot the $n$th-order R{\'e}nyi entropy $S_A^{(n)}$ as a function of $n$ together with the von Neumann entropy $S_A$ at $n=1$. Note that $S_A^{(1/2)}$ is obtained by the square root of the density matrix. We observe that $S_A^{(n)}$ rapidly increases toward $n\rightarrow 0$ in the region of $n\simlt 3$. As a result, neither $S_A^{(2)}$ nor $S_A^{(1/2)}$ is a good approximation to the von Neumann entropy. Futhermore, this makes the precise extrapolation of  $S_A^{(2)}$ with $(n\ge 2)$ difficult as shown with blue and green broken lines in Fig.~\ref{fig:renyi_n}.

\subsection{Quantum phase transition with $\mu\ne 0$}
\label{subsec:qpt}

We evaluate the number density with the numerical differentiation of $f$:
\be
\langle n\rangle=\frac{\partial}{\partial \mu}f\approx \frac{f(\mu+\Delta\mu)-f(\mu-\Delta\mu)}{2\Delta\mu}= \frac{-1}{L N_t}\frac{\ln Z(\mu+\Delta\mu)-\ln Z(\mu-\Delta\mu)}{2\Delta\mu},
\ee
where the partition function $Z$ is evaluated with the HOTRG algorithm with the  bond dimensions $D_{\rm cut}=125$, 130 and 135. We focus on $\beta=1.4$ in the $\mu\ne 0$ case.

At the criticality of the second order phase tranasition the spatial correlation length $\xi$ should diverge as $\xi\sim \delta^{-\nu}=\vert \mu-\mu_{\rm c}\vert^{-\nu}$  with  $\nu$ the critical exponent. Since our model defined in Eq.~(\ref{eq:action}) breakes the space-time symmetry due to the introduction of the chemical potential, the temporal correlation length $\xi_t$ should be deviated from $\xi$ and both are related by $\xi_t\sim \xi^z\sim \delta^{-z\nu}$ with $z$ the dynamical critical exponent. 

\begin{figure}[htbp]
	\begin{minipage}[t]{0.48\hsize}
    		\centering
	\includegraphics[width=1.0\hsize]{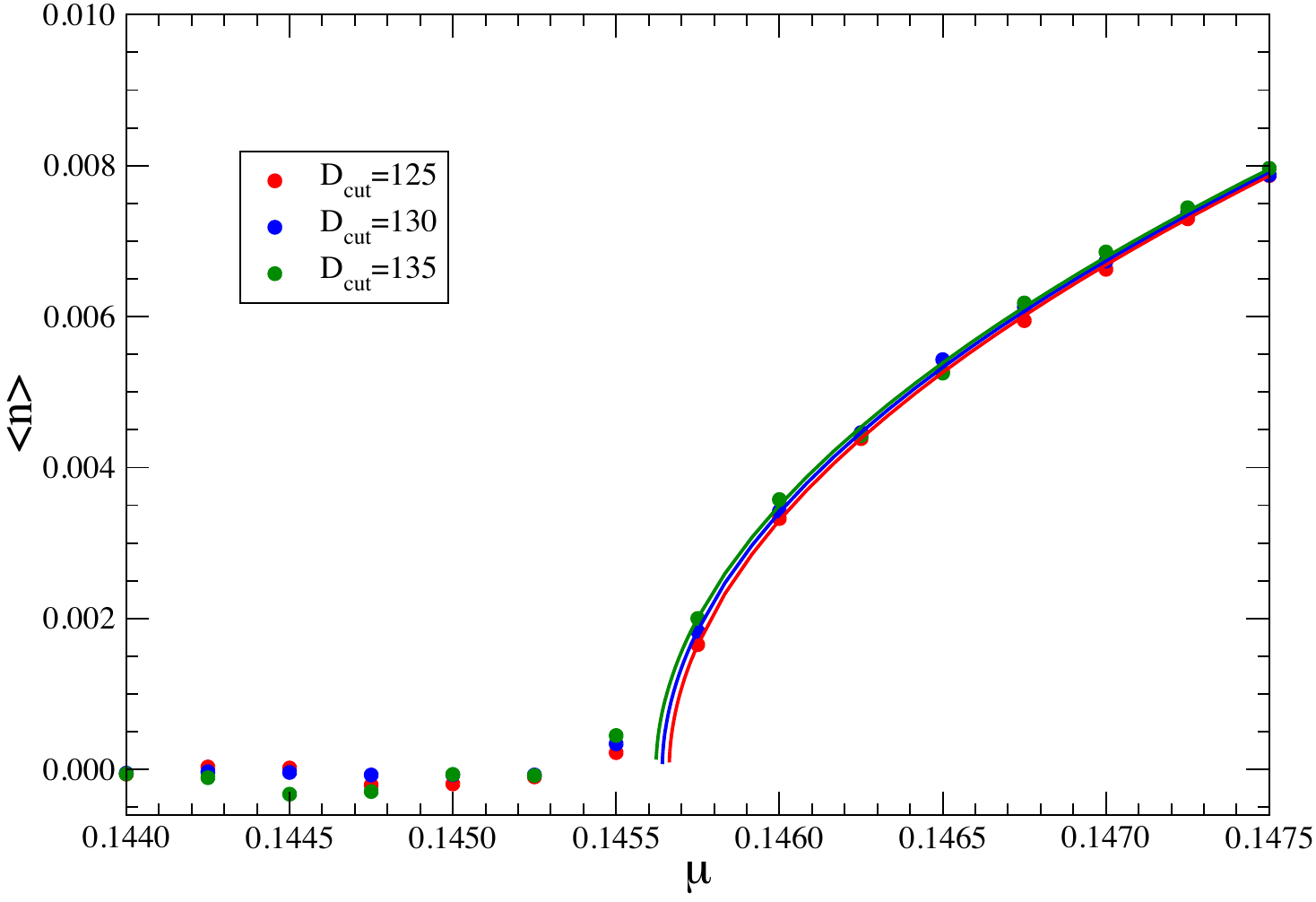}
	\caption{$\mu$ dependence of number density $\langle n\rangle$ at $\beta=1.4$ on a $2^{25}\times 2^{25}$ lattice with $D_{\rm cut}\in[125,135]$. The solid curves represent the fit results at $D_{\rm cut}=125$(red), 130(blue) and 135(green).}
  	\label{fig:n}
  	\end{minipage}
  	\hspace*{3mm}
	\begin{minipage}[t]{0.48\hsize}
    		\centering
 	\includegraphics[width=0.9\hsize]{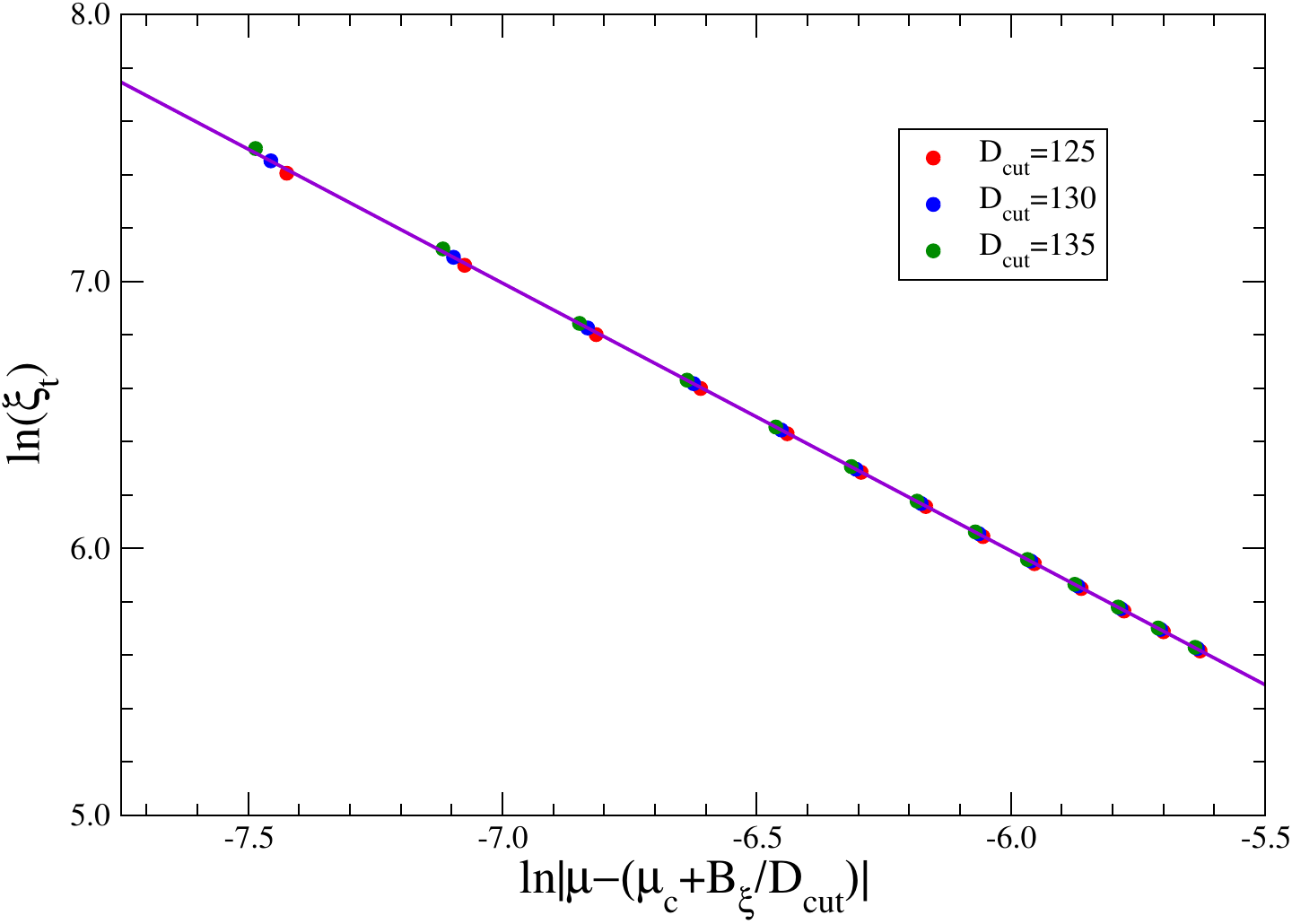}
        \caption{Temporal correlation length $\ln(\xi_t)$ at $\beta=1.4$ as a function of $\ln\vert \mu-(\mu_{\rm c}+B_\xi/D_{\rm cut})\vert$ with $D_{\rm cut}\in[125,135]$. The solid curve represents the fit result choosing $D_{\rm cut}=\infty$.}
  	\label{fig:xi_t_mu}
	\end{minipage}
\end{figure}

Figure~\ref{fig:n} plots the number density as a function of $\mu$ around the transition point on a $V=L\times N_t=2^{25}\times 2^{25}$ lattice with $D_{\rm cut}\in [125,135]$. The volume is large enough to be regarded as the thermodynamic limit at zero temperature: $T/m=2.1\times 10^{-7}$ and $Lm=4.9\times 10^{6}$ with the mass gap $m$. Taking account of the slight $D_{\rm cut}$ dependence we apply the global fit to the data in the range of $0.14575\le \mu\le 0.14700$ at $D_{\rm cut}\in [125,135]$ assuming the function form of $\langle n\rangle (\mu,D_{\rm cut})=A_n\cdot \left\{\mu-(\mu_{\rm c}+B_n/D_{\rm cut})\right\}^\nu$ with $A_n$, $\mu_{\rm c}$, $B_n$ and $\nu$ the fit parameters. The solid curves show the fit results with $A_n=0.20(2)$, $\mu_{\rm c}=0.14512(11)$, $B_n=0.068(12)$ and $\nu=0.512(15)$, where the value of $\mu_{\rm c}$ is consistent with the mass gap $m=1/\xi_0=1/6.90(1)=0.1449(2)$ at $\mu=0$ obtained by a high precision Monte Carlo result~\cite{Wolff:1989hv}. 

In Fig.~\ref{fig:xi_t_mu} we show the results for the global fit of the temporal correlation length at $D_{\rm cut}\in [125,135]$ employing the fit form of $\ln \xi_t(\mu,D_{\rm cut})=A_\xi+\alpha \ln|\mu-(\mu_{\rm c}+B_\xi/D_{\rm cut})|$ with $\mu_{\rm c}=0.14512$.  The solid curve, which shows fairly linear behavior, is drawn with the fit results of $A_\xi=-0.030(29)$, $B_\xi=0.0599(9)$ and $\alpha=1.003(5)$ choosing $D_{\rm cut}=\infty$. The relation $\alpha=z\nu$ with the use of $\nu=0.512(15)$ gives the dynamical critical exponent $z=1.96(6)$. Our results of $\nu=0.512(15)$ and $z=1.96(6)$ are consistent with the theoretical expectation of $\nu=0.5$ and $z=2$.

\section{Summary} 
\label{sec:summary}

We have studied the (1+1)$d$ O(3) NLSM with the TRG method. At $\mu=0$ we calcuate the von Neumann and R\'enyi types of entanglement entropies. The central charge obtained from the asymptotic scaling behavior of the von Neumann entropy is $c=1.97(9)$, which is consistent with $c=2.04(14)$ previously obtained with the MPS method. The direct comparison between both entropies implies that it may be difficult to estimate the von Neumann entropy at high precision from the extrapolation of higher-order R\'enyi etropies. We have also investigated the properties of the quantum phase transition with finite $\mu$, which causes the sign problem in the Monte Carlo approach. We find the critical chemical potential $\mu_{\rm c}=0.14512(11)$ at $\beta=1.4$ in the limit of $D_{\rm cut}\rightarrow \infty$, which is consistent with the mass gap $m=1/\xi_0=1/6.90(1)=0.1449(2)$ at $\mu=0$ obtained in the Monte Carlo approach~\cite{Wolff:1989hv}. Our results for the critical exponent $\nu=0.512(15)$ and the dynamical one $z=1.96(6)$ also show consistency with the theoretical expectation of $\nu=0.5$ and $z=2$. This is the first successful calculation of the dynamical critical exponent with the TRG method.

\begin{acknowledgments}
Numerical calculation for the present work was carried out with the supercomputers Cygnus and Pegasus under the Multidisciplinary Cooperative Research Program of Center for Computational Sciences, University of Tsukuba. We also used the supercomputer Fugaku provided by RIKEN through the HPCI System Research Project (Project ID: hp220203, hp230247).
This work is supported in part by Grants-in-Aid for Scientific Research from the Ministry of Education, Culture, Sports, Science and Technology (MEXT) (Nos. 24H00214, 24H00940).

\end{acknowledgments}

\bibliographystyle{JHEP}
\bibliography{bib/formulation,bib/algorithm,bib/discrete,bib/grassmann,bib/continuous,bib/gauge,bib/review,bib/for_this_paper}

\providecommand{\href}[2]{#2}\begingroup\raggedright\begin{thebibliography}{10}

\bibitem{Levin:2006jai}
M.~Levin and C.~P. Nave, \emph{{Tensor renormalization group approach to
  two-dimensional classical lattice models}},
  \href{https://doi.org/10.1103/PhysRevLett.99.120601}{\emph{Phys. Rev. Lett.}
  {\bfseries 99} (2007) 120601},
  [\href{https://arxiv.org/abs/cond-mat/0611687}{{\ttfamily
  cond-mat/0611687}}].

\bibitem{PhysRevB.86.045139}
Z.~Y. Xie, J.~Chen, M.~P. Qin, J.~W. Zhu, L.~P. Yang and T.~Xiang,
  \emph{Coarse-graining renormalization by higher-order singular value
  decomposition}, \href{https://doi.org/10.1103/PhysRevB.86.045139}{\emph{Phys.
  Rev. B} {\bfseries 86} (2012) 045139},
  [\href{https://arxiv.org/abs/1201.1144}{{\ttfamily 1201.1144}}].

\bibitem{Shimizu:2014uva}
Y.~Shimizu and Y.~Kuramashi, \emph{{Grassmann tensor renormalization group
  approach to one-flavor lattice Schwinger model}},
  \href{https://doi.org/10.1103/PhysRevD.90.014508}{\emph{Phys. Rev.}
  {\bfseries D90} (2014) 014508},
  [\href{https://arxiv.org/abs/1403.0642}{{\ttfamily 1403.0642}}].

\bibitem{PhysRevLett.115.180405}
G.~Evenbly and G.~Vidal, \emph{Tensor network renormalization},
  \href{https://doi.org/10.1103/PhysRevLett.115.180405}{\emph{Phys. Rev. Lett.}
  {\bfseries 115} (2015) 180405}.

\bibitem{Sakai:2017jwp}
R.~Sakai, S.~Takeda and Y.~Yoshimura, \emph{{Higher order tensor
  renormalization group for relativistic fermion systems}},
  \href{https://doi.org/10.1093/ptep/ptx080}{\emph{PTEP} {\bfseries 2017}
  (2017) 063B07}, [\href{https://arxiv.org/abs/1705.07764}{{\ttfamily
  1705.07764}}].

\bibitem{PhysRevLett.118.110504}
S.~Yang, Z.-C. Gu and X.-G. Wen, \emph{Loop optimization for tensor network
  renormalization},
  \href{https://doi.org/10.1103/PhysRevLett.118.110504}{\emph{Phys. Rev. Lett.}
  {\bfseries 118} (2017) 110504}.

\bibitem{Hauru:2017tne}
M.~Hauru, C.~Delcamp and S.~Mizera, \emph{{Renormalization of tensor networks
  using graph independent local truncations}},
  \href{https://doi.org/10.1103/PhysRevB.97.045111}{\emph{Phys. Rev.}
  {\bfseries B97} (2018) 045111},
  [\href{https://arxiv.org/abs/1709.07460}{{\ttfamily 1709.07460}}].

\bibitem{Adachi:2019paf}
D.~Adachi, T.~Okubo and S.~Todo, \emph{{Anisotropic Tensor Renormalization
  Group}}, \href{https://doi.org/10.1103/PhysRevB.102.054432}{\emph{Phys. Rev.
  B} {\bfseries 102} (2020) 054432},
  [\href{https://arxiv.org/abs/1906.02007}{{\ttfamily 1906.02007}}].

\bibitem{Kadoh:2019kqk}
D.~Kadoh and K.~Nakayama, \emph{{Renormalization group on a triad network}},
  \href{https://arxiv.org/abs/1912.02414}{{\ttfamily 1912.02414}}.

\bibitem{Akiyama:2020soe}
S.~Akiyama, Y.~Kuramashi, T.~Yamashita and Y.~Yoshimura, \emph{{Restoration of
  chiral symmetry in cold and dense Nambu--Jona-Lasinio model with tensor
  renormalization group}},
  \href{https://doi.org/10.1007/JHEP01(2021)121}{\emph{JHEP} {\bfseries 01}
  (2021) 121}, [\href{https://arxiv.org/abs/2009.11583}{{\ttfamily
  2009.11583}}].

\bibitem{PhysRevB.105.L060402}
D.~Adachi, T.~Okubo and S.~Todo, \emph{Bond-weighted tensor renormalization
  group}, \href{https://doi.org/10.1103/PhysRevB.105.L060402}{\emph{Phys. Rev.
  B} {\bfseries 105} (2022) L060402},
  [\href{https://arxiv.org/abs/2011.01679}{{\ttfamily 2011.01679}}].

\bibitem{Akiyama:2022pse}
S.~Akiyama, \emph{{Bond-weighting method for the Grassmann tensor
  renormalization group}},
  \href{https://doi.org/10.1007/JHEP11(2022)030}{\emph{JHEP} {\bfseries 11}
  (2022) 030}, [\href{https://arxiv.org/abs/2208.03227}{{\ttfamily
  2208.03227}}].

\bibitem{Akiyama:2019xzy}
S.~Akiyama, Y.~Kuramashi, T.~Yamashita and Y.~Yoshimura, \emph{{Phase
  transition of four-dimensional Ising model with higher-order tensor
  renormalization group}},
  \href{https://doi.org/10.1103/PhysRevD.100.054510}{\emph{Phys. Rev.}
  {\bfseries D100} (2019) 054510},
  [\href{https://arxiv.org/abs/1906.06060}{{\ttfamily 1906.06060}}].

\bibitem{Akiyama:2020ntf}
S.~Akiyama, D.~Kadoh, Y.~Kuramashi, T.~Yamashita and Y.~Yoshimura,
  \emph{{Tensor renormalization group approach to four-dimensional complex
  $\phi^4$ theory at finite density}},
  \href{https://doi.org/10.1007/JHEP09(2020)177}{\emph{JHEP} {\bfseries 09}
  (2020) 177}, [\href{https://arxiv.org/abs/2005.04645}{{\ttfamily
  2005.04645}}].

\bibitem{Akiyama:2021zhf}
S.~Akiyama, Y.~Kuramashi and Y.~Yoshimura, \emph{{Phase transition of
  four-dimensional lattice $\phi^4$ theory with tensor renormalization group}},
  \href{https://doi.org/10.1103/PhysRevD.104.034507}{\emph{Phys. Rev. D}
  {\bfseries 104} (2021) 034507},
  [\href{https://arxiv.org/abs/2101.06953}{{\ttfamily 2101.06953}}].

\bibitem{Akiyama:2022eip}
S.~Akiyama and Y.~Kuramashi, \emph{{Tensor renormalization group study of
  (3+1)-dimensional \ensuremath{\mathbb{Z}}$_{2}$ gauge-Higgs model at finite
  density}}, \href{https://doi.org/10.1007/JHEP05(2022)102}{\emph{JHEP}
  {\bfseries 05} (2022) 102},
  [\href{https://arxiv.org/abs/2202.10051}{{\ttfamily 2202.10051}}].

\bibitem{Akiyama:2023hvt}
S.~Akiyama and Y.~Kuramashi, \emph{{Critical endpoint of (3+1)-dimensional
  finite density \ensuremath{\mathbb{Z}}$_{3}$ gauge-Higgs model with tensor
  renormalization group}},
  \href{https://doi.org/10.1007/JHEP10(2023)077}{\emph{JHEP} {\bfseries 10}
  (2023) 077}, [\href{https://arxiv.org/abs/2304.07934}{{\ttfamily
  2304.07934}}].

\bibitem{Shimizu:2014fsa}
Y.~Shimizu and Y.~Kuramashi, \emph{{Critical behavior of the lattice Schwinger
  model with a topological term at $\theta=\pi$ using the Grassmann tensor
  renormalization group}},
  \href{https://doi.org/10.1103/PhysRevD.90.074503}{\emph{Phys. Rev.}
  {\bfseries D90} (2014) 074503},
  [\href{https://arxiv.org/abs/1408.0897}{{\ttfamily 1408.0897}}].

\bibitem{Kawauchi:2016xng}
H.~Kawauchi and S.~Takeda, \emph{{Tensor renormalization group analysis of
  CP($N$-1) model}},
  \href{https://doi.org/10.1103/PhysRevD.93.114503}{\emph{Phys. Rev.}
  {\bfseries D93} (2016) 114503},
  [\href{https://arxiv.org/abs/1603.09455}{{\ttfamily 1603.09455}}].

\bibitem{Kawauchi:2016dcg}
H.~Kawauchi and S.~Takeda, \emph{{Phase structure analysis of CP(N-1) model
  using Tensor renormalization group}},
  \href{https://doi.org/10.22323/1.256.0322}{\emph{PoS} {\bfseries LATTICE2016}
  (2016) 322}, [\href{https://arxiv.org/abs/1611.00921}{{\ttfamily
  1611.00921}}].

\bibitem{Yang:2015rra}
L.-P. Yang, Y.~Liu, H.~Zou, Z.~Xie and Y.~Meurice, \emph{{Fine structure of the
  entanglement entropy in the O(2) model}},
  \href{https://doi.org/10.1103/PhysRevE.93.012138}{\emph{Phys. Rev. E}
  {\bfseries 93} (2016) 012138},
  [\href{https://arxiv.org/abs/1507.01471}{{\ttfamily 1507.01471}}].

\bibitem{Shimizu:2017onf}
Y.~Shimizu and Y.~Kuramashi, \emph{{Berezinskii-Kosterlitz-Thouless transition
  in lattice Schwinger model with one flavor of Wilson fermion}},
  \href{https://doi.org/10.1103/PhysRevD.97.034502}{\emph{Phys. Rev.}
  {\bfseries D97} (2018) 034502},
  [\href{https://arxiv.org/abs/1712.07808}{{\ttfamily 1712.07808}}].

\bibitem{Takeda:2014vwa}
S.~Takeda and Y.~Yoshimura, \emph{{Grassmann tensor renormalization group for
  the one-flavor lattice Gross-Neveu model with finite chemical potential}},
  \href{https://doi.org/10.1093/ptep/ptv022}{\emph{PTEP} {\bfseries 2015}
  (2015) 043B01}, [\href{https://arxiv.org/abs/1412.7855}{{\ttfamily
  1412.7855}}].

\bibitem{Kadoh:2018hqq}
D.~Kadoh, Y.~Kuramashi, Y.~Nakamura, R.~Sakai, S.~Takeda and Y.~Yoshimura,
  \emph{{Tensor network formulation for two-dimensional lattice $ \mathcal{N} $
  = 1 Wess-Zumino model}},
  \href{https://doi.org/10.1007/JHEP03(2018)141}{\emph{JHEP} {\bfseries 03}
  (2018) 141}, [\href{https://arxiv.org/abs/1801.04183}{{\ttfamily
  1801.04183}}].

\bibitem{Kadoh:2019ube}
D.~Kadoh, Y.~Kuramashi, Y.~Nakamura, R.~Sakai, S.~Takeda and Y.~Yoshimura,
  \emph{{Investigation of complex $\phi^{4}$ theory at finite density in two
  dimensions using TRG}},
  \href{https://doi.org/10.1007/JHEP02(2020)161}{\emph{JHEP} {\bfseries 02}
  (2020) 161}, [\href{https://arxiv.org/abs/1912.13092}{{\ttfamily
  1912.13092}}].

\bibitem{Kuramashi:2019cgs}
Y.~Kuramashi and Y.~Yoshimura, \emph{{Tensor renormalization group study of
  two-dimensional U(1) lattice gauge theory with a $\theta$ term}},
  \href{https://doi.org/10.1007/JHEP04(2020)089}{\emph{JHEP} {\bfseries 04}
  (2020) 089}, [\href{https://arxiv.org/abs/1911.06480}{{\ttfamily
  1911.06480}}].

\bibitem{Akiyama:2021xxr}
S.~Akiyama and Y.~Kuramashi, \emph{{Tensor renormalization group approach to
  (1+1)-dimensional Hubbard model}},
  \href{https://doi.org/10.1103/PhysRevD.104.014504}{\emph{Phys. Rev. D}
  {\bfseries 104} (2021) 014504},
  [\href{https://arxiv.org/abs/2105.00372}{{\ttfamily 2105.00372}}].

\bibitem{Akiyama:2021glo}
S.~Akiyama, Y.~Kuramashi and T.~Yamashita, \emph{{Metal\textendash{}insulator
  transition in the (2+1)-dimensional Hubbard model with the tensor
  renormalization group}},
  \href{https://doi.org/10.1093/ptep/ptac014}{\emph{PTEP} {\bfseries 2022}
  (2022) 023I01}, [\href{https://arxiv.org/abs/2109.14149}{{\ttfamily
  2109.14149}}].

\bibitem{Nakayama:2021iyp}
K.~Nakayama, L.~Funcke, K.~Jansen, Y.-J. Kao and S.~K\"uhn, \emph{{Phase
  structure of the CP(1) model in the presence of a topological
  \ensuremath{\theta}-term}},
  \href{https://doi.org/10.1103/PhysRevD.105.054507}{\emph{Phys. Rev. D}
  {\bfseries 105} (2022) 054507},
  [\href{https://arxiv.org/abs/2107.14220}{{\ttfamily 2107.14220}}].

\bibitem{Luo:2022eje}
X.~Luo and Y.~Kuramashi, \emph{{Tensor renormalization group approach to
  (1+1)-dimensional SU(2) principal chiral model at finite density}},
  \href{https://doi.org/10.1103/PhysRevD.107.094509}{\emph{Phys. Rev. D}
  {\bfseries 107} (2023) 094509},
  [\href{https://arxiv.org/abs/2208.13991}{{\ttfamily 2208.13991}}].

\bibitem{Akiyama:2024qer}
S.~Akiyama and Y.~Kuramashi, \emph{{Tensor renormalization group study of (1 +
  1)-dimensional U(1) gauge-Higgs model at \ensuremath{\theta} =
  \ensuremath{\pi} with L\"uscher\textquoteright{}s admissibility condition}},
  \href{https://doi.org/10.1007/JHEP09(2024)086}{\emph{JHEP} {\bfseries 09}
  (2024) 086}, [\href{https://arxiv.org/abs/2407.10409}{{\ttfamily
  2407.10409}}].

\bibitem{Luo:2023ont}
X.~Luo and Y.~Kuramashi, \emph{{Entanglement and R\'enyi entropies of
  (1+1)-dimensional O(3) nonlinear sigma model with tensor renormalization
  group}}, \href{https://doi.org/10.1007/JHEP03(2024)020}{\emph{JHEP}
  {\bfseries 03} (2024) 020},
  [\href{https://arxiv.org/abs/2308.02798}{{\ttfamily 2308.02798}}].

\bibitem{Luo:2024lbh}
X.~Luo and Y.~Kuramashi, \emph{{Quantum phase transition of (1+1)-dimensional
  O(3) nonlinear sigma model at finite density with tensor renormalization
  group}}, \href{https://doi.org/10.1007/JHEP11(2024)144}{\emph{JHEP}
  {\bfseries 11} (2024) 144},
  [\href{https://arxiv.org/abs/2406.08865}{{\ttfamily 2406.08865}}].

\bibitem{Dzhaparidze_1978}
G.~I. Dzhaparidze and A.~A. Nersesyan, \emph{Magnetic-field phase transition in
  a one-dimensional system of electrons with attraction}, {\emph{JETP Lett.}
  {\bfseries 27} (1978) 356}.

\bibitem{Pokrovsky_1979}
V.~L. Pokrovsky and A.~L. Talapov, \emph{Ground state, spectrum, and phase
  diagram of two-dimensional incommensurate crystals},
  \href{https://doi.org/10.1103/PhysRevLett.42.65}{\emph{Phys. Rev. Lett.}
  {\bfseries 42} (1979) 65}.

\bibitem{Schulz_1980}
H.~J. Schulz, \emph{Critical behavior of commensurate-incommensurate phase
  transitions in two dimensions},
  \href{https://doi.org/10.1103/PhysRevB.22.5274}{\emph{Phys. Rev. B}
  {\bfseries 22} (1980) 5274}.

\bibitem{Schulz_1986}
H.~J. Schulz, \emph{Phase diagrams and correlation exponents for quantum spin
  chains of arbitrary spin quantum number},
  \href{https://doi.org/10.1103/PhysRevB.34.6372}{\emph{Phys. Rev. B}
  {\bfseries 34} (1986) 6372}.

\bibitem{Affleck_1990}
I.~Affleck, \emph{Theory of haldane-gap antiferromagnets in applied fields},
  \href{https://doi.org/10.1103/PhysRevB.41.6697}{\emph{Phys. Rev. B}
  {\bfseries 41} (1990) 6697}.

\bibitem{Wolff:1989hv}
U.~Wolff, \emph{{Asymptotic Freedom and Mass Generation in the O(3) Nonlinear
  $\sigma$ Model}},
  \href{https://doi.org/10.1016/0550-3213(90)90313-3}{\emph{Nucl. Phys. B}
  {\bfseries 334} 581}.

\bibitem{Calabrese:2004eu}
P.~Calabrese and J.~L. Cardy, \emph{{Entanglement entropy and quantum field
  theory}}, \href{https://doi.org/10.1088/1742-5468/2004/06/P06002}{\emph{J.
  Stat. Mech.} {\bfseries 0406} (2004) P06002},
  [\href{https://arxiv.org/abs/hep-th/0405152}{{\ttfamily hep-th/0405152}}].

\bibitem{Bruckmann:2018usp}
F.~Bruckmann, K.~Jansen and S.~K\"uhn, \emph{{O(3) nonlinear sigma model in 1+1
  dimensions with matrix product states}},
  \href{https://doi.org/10.1103/PhysRevD.99.074501}{\emph{Phys. Rev. D}
  {\bfseries 99} (2019) 074501},
  [\href{https://arxiv.org/abs/1812.00944}{{\ttfamily 1812.00944}}].

\end{thebibliography}\endgroup

%

\end{document}